Article

# Mechanical transistors for logic-with-memory computing

Huyue Chen[1,2], Chao Song[1], Jiahao Wu[1], Bihui Zou[1], Zhihan Zhang[1], An Zou[1], Yuljae Cho[1], Zhaoguang Wang[1], Wenming Zhang[2], Lei Shao[1,2]* & Jaehyung Ju[1,*]

## Abstract

As a potential revolutionary topic in future information processing, mechanical computing[1,2] has gained tremendous attention for replacing or supplementing conventional electronics vulnerable to power outages, security attacks, and harsh environments[3,4]. Despite its potential for constructing intelligent matter towards nonclassical computing systems beyond the von Neumann architecture, most works on mechanical computing demonstrated that the ad hoc design of simple logic gates cannot fully realize a universal mechanical processing framework involving interconnected arithmetic logic components and memory[5-16]. However, such a logic-with-memory computing architecture is critical for complex and persistent state-dependent computations such as sequential logic. Here we propose a mechanical transistor (M-Transistor), abstracting omnipresent temperatures as the input-output mechanical bits, which consists of a metamaterial thermal channel as the gate terminal driving a nonlinear bistable soft actuator to selectively connect the output terminal to two other variable thermal sources. This M-Transistor is an elementary unit to modularly form various combinational and sequential circuits, such as complex logic gates, registers (volatile memory), and long-term memories (non-volatile memory) with much fewer units than the electronic counterparts. Moreover, they can establish a universal processing core comprising an arithmetic circuit and a register in a compact, reprogrammable network involving periodic read, write, memory, and logic operations of the mechanical bits. Additionally, we demonstrate the self-unfolding of aerospace solar sails deployed by sequential logic functions with environmental thermal inputs. Our work contributes to realizing a non-electric universal mechanical computing architecture that combines

1 University of Michigan - Shanghai Jiao Tong University Joint Institute, Shanghai Jiao Tong University, Shanghai, P. R. China.
2 School of Mechanical Engineering, Shanghai Jiao Tong University, Shanghai, P. R. China.
* e-mail: L. S; J. J. (for arXiv: H.C.: huyue_chen@sjtu.edu.cn)

multidisciplinary engineering with structural mechanics, materials science, thermal engineering, physical intelligence, and computational science.

## Introduction

Electrical computing has been indispensable since the invention of transistors[17,18]. However, electronics fail in harsh environments, such as extreme temperatures and radiation[3,4], and they cannot be engineered to directly interact with external stimuli such as heat, force, pressure, etc. Recently, mechanical computing has been gaining attention as a new computational paradigm to supplement or replace conventional computing[1,2]. Unlike the earlier obsolete analog mechanical computing engines[19,20], modern mechanical logic gates integrate abstracted digital processing[21-23], which can increase the computational density and respond to environmental stimuli. For instance, several mechanical devices were explored to implement binary information processes with mechanical metamaterials[5-8,21], specific solvents[9,10], pneumatic[11,12], magnetic[13,14], and thermal control[15,16]. However, most previous studies only covered basic logic computing or memory bits[24-27] separately, without forming integrated architectures to demonstrate a full processing core with key features, such as computing, transferring, and registering mechanical bit signals. This restriction is due to the ad hoc design of the above building blocks, causing an intractable barrier between logic and memory, limiting their development towards higher levels of physical intelligence and distributed information processing power[28-30]. Additionally, some metamaterial-based computing systems heavily rely on manual resetting or electrical signals for either inputs or outputs[31-35]. This results in a heterogeneous signal flow and inconvenience in logic-memory integration. Therefore, a more fundamental approach is required, wherein an elementary mechanical transistor is proposed to construct interconnected logic and memory components that perform persistent computation and complex decision-making.

Inspired by the omnipresence of natural heat as a mechanical signal, we created a mechanical transistor with four thermal terminals driven by and output heat. This transistor, which abstracts high or low temperatures as mechanical 1 or 0, respectively, comprises a mechanical metamaterial temperature-sensing channel that actuates a bistable soft element to connect the output terminal from two other variable thermal sources. This unconventional transistor design, with one more source terminal,

**Article**

significantly promotes reprogrammability and reduces the number of components required to construct various logic and memory units, which is especially desirable for resource-constrained scenarios. Thus, we used far fewer transistors to construct a mechanical logic-with-memory processing core. This shows the multicycle computing of adding two 4-bit numbers in an arithmetic logic unit while continuously memorizing and reading from an on-site register. This fully mechanical logic-with-memory computing framework demonstrates the potential towards and beyond the von Neumann architecture and could be used for high-level autonomous matter and physical intelligent systems.

# Article

# Mechanical transistors

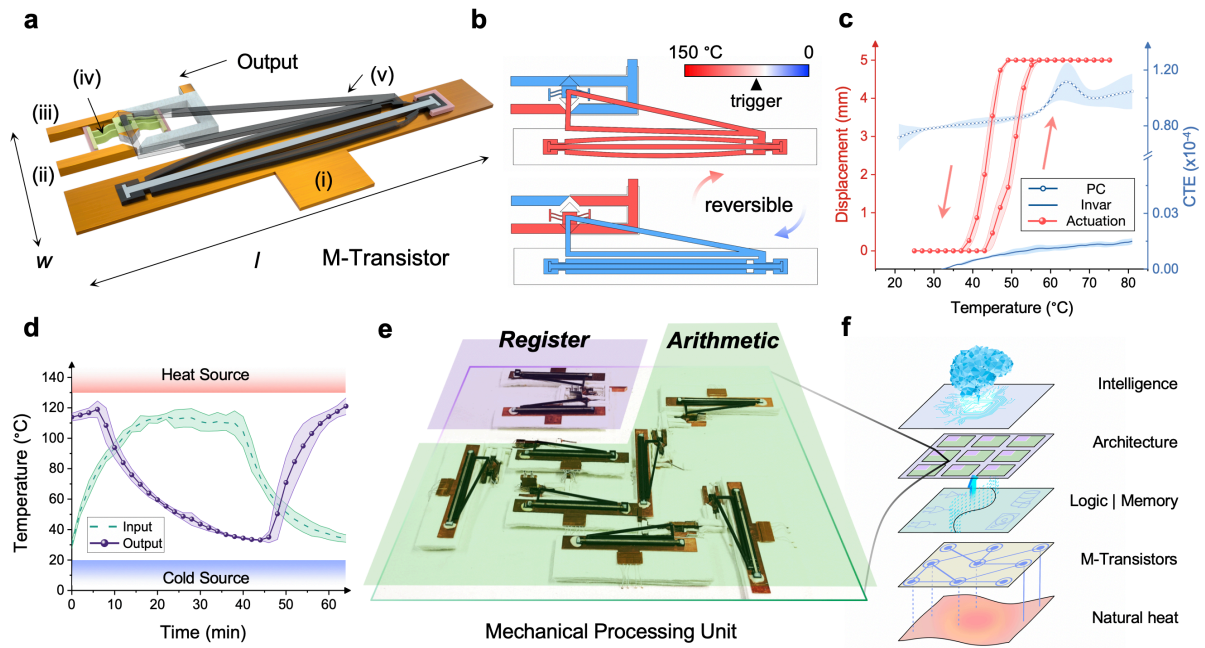

**Fig.1|Mechanical transistor and logic-with-memory computing architecture. a,** A 3D schematic of a mechanical transistor (M-Transistor) comprising (i)-(iii) three input terminals and one output terminal for carrying temperature signals, (iv) a bistable soft actuator, and (v) a mechanical metamaterial sensor. The value of *l* and *w* are 250 mm and 85 mm, respectively. **b,** Reversible switching mechanism between the two thermal conditions as the mechanical sensor expands when sensing a high temperature and shrinks when exposed to a low temperature. **c,** Snap-through and snap-back actuation in experiments. The blue lines are the coefficient of thermal expansion (CTE) of polycarbonate (PC) and invar alloy; while the red curves represent reversible bistability within a preset displacement (5mm). **d,** Thermal response of an M-Transistor as its terminals set up in **b**. High (abstracted as 1, >60°C) and low (abstracted as 0, <40°C) output temperature states exhibit a quantitative and significant distinction, thus facilitating the binary abstraction. **e,** The mechanical logic-with-memory processing unit network, with an integrated arithmetic logic unit and an on-site register memory in one core. **f,** Five-level hierarchy of mechanical computing, showing that M-Transistors eliminate the barriers between logic and memory, promoting higher-level architectural integrations.

As the building block of a non-electric computer, we create an elementary mechanical transistor (Fig. 1a), instead of directly starting with logic-level designs, to modularly construct fully functional logic circuits and with-memory processing networks. The M-Transistor consists of three thermal inputs (i)-(iii) and one thermal output, a bistable soft

# Article

actuator (iv), and a flexible metamaterial sensor (v) (details in "Materials and Methods"). In contrast to previous manual operations of mechanical logic devices[5-8,21,31-33], we encode natural heat into the M-Transistor, abstracting "1" for temperatures above 60 °C (red) and "0" for below 40 °C (blue), to self-switch between states due to thermomechanical actuation. Depending on the adaptive expansion or shrinkage of the metamaterial sensor with respect to environmental temperatures (input i), inputs (ii) or (iii) are selected to connect with the output channel through a triangular copper block mounted at the end of a soft bistable element (Fig. 1b). This copper maintains conductivity in one of the thermal channels, while disconnecting the other channel. Depending on the various setup of sources (ii) and (iii), we could implement diverse logic and memory functions in a single M-Transistor.

The mechanical sensor yields a large displacement triggered by a thermal expansion mismatch between a polycarbonate Kirigami-inspired displacement amplifier and an I-shape invar bar[35-38]. We further maximize the vertical displacement by breaking the mirror symmetry about the vertical centerline of the kirigami structure (details in Supplementary Materials "Thermomechanical deformation of Compliant Porous Structure (CPS)"). Therefore, as the sensor temperature increases from room temperature (25 °C) to 80 °C, the bistable actuator snaps through and converts the continuous signal of the metamaterial sensor into a distinct binary, while the 'trigger point' is between 50 and 60°C (Fig. 1c). To avoid ambiguity in the transistor operation, the pre-designed asymmetric bending curves in the soft beams prescribes the bistable elements to reside at one side in the initial cold state (details in Supplementary Materials "Bistable modeling"). Because the sensor's thermomechanical forces at high temperatures ($T_H$) and low temperatures ($T_L$) are designed to overcome the bistable energy barriers, the snap-through/back features regulate the reversibility of M-Transistors. We verify the M-Transistors by recording the thermal signals at the T-shaped copper (i) and the output with constant heating (>130 °C) and cooling (< 30 °C) sources at inputs (ii) and (iii), respectively. The output thermal signal responds in the opposite manner to the input, as shown in Fig. 1d (details in Supplementary Video 2 "Thermomechanical actuating of the transistor").

# Article

Notably, due to the distinction between logic and memory, scientists have failed to create information interactions between mechanical arithmetic[19-22,31-34] and storage[24-27]. Herein, we demonstrate a mechanical computing core integrated with an arithmetic logic unit (ALU) and a register (Reg) based on M-Transistors (Fig. 1e). Future mechanical computing expects a neuromorphic morphology, instead of the von Neumann architecture which separates logic and memory (Fig. 1f). A significant advantage of logic-with-memory computing is that we can save the required information after one computation cycle, and subsequently replenish the recorded data. M-Transistors, as a universal unit, could be used to demonstrate all logic gates, rewritable registers, non-volatile memories, full adders, and other physical intelligence architectures (details in Supplementary Video 1 "Mechanical logic-with-memory architecture").

# Article

# Mechanical logic gates

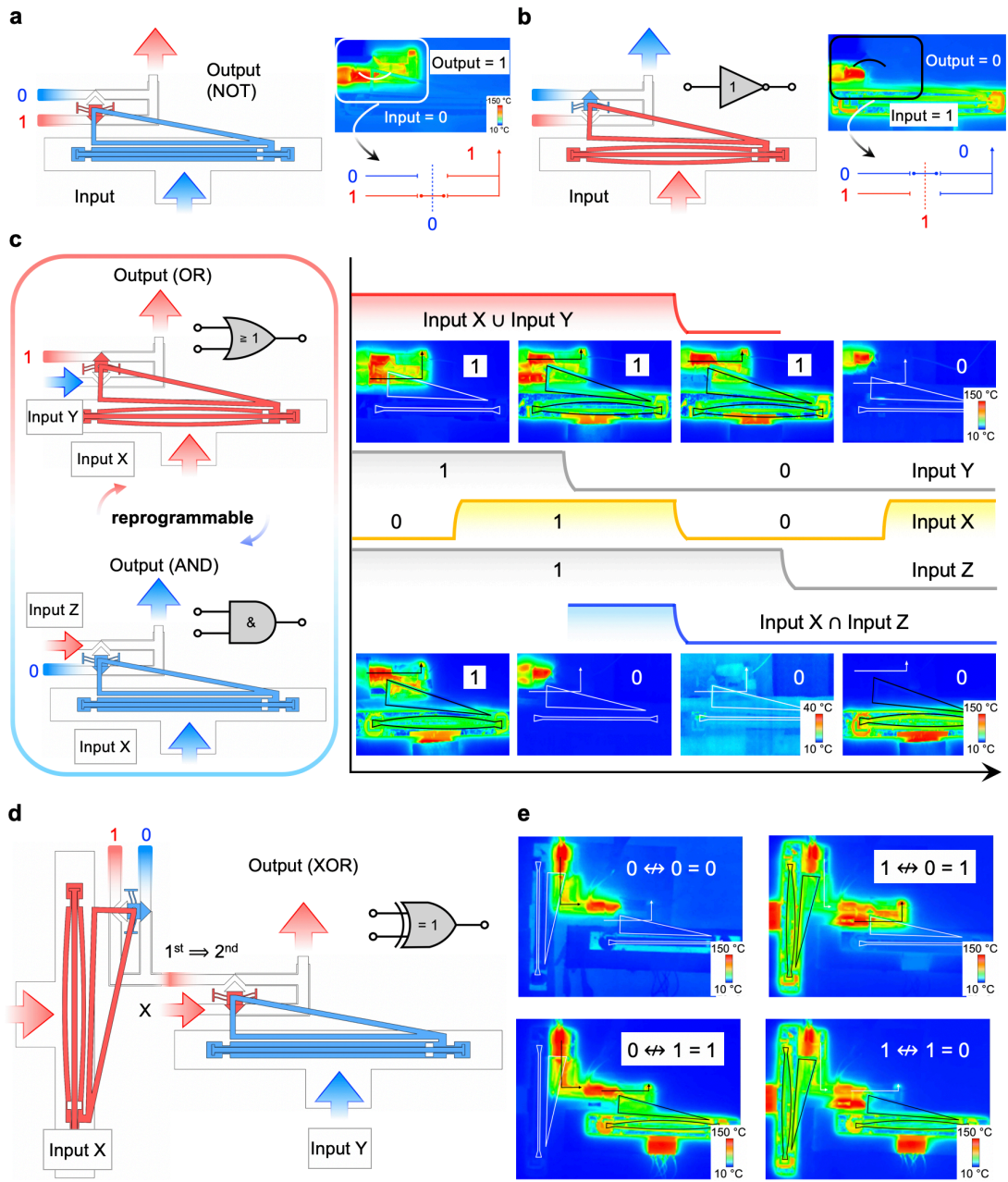

**Fig.2|Reprogrammable mechanical logic gates with complete functions. a,b,** NOT gate yielding an output opposite to input (i), with constant heat (ii) and cold (iii) sources. Infrared images and heat path diagrams describe two reversible states. Red represents a high temperature as a bit value of 1, and blue denotes a low temperature as a bit value of 0. **c,** A Reprogrammable strategy utilizing the identical physical configuration. An OR gate calculates inputs X and Y, and an AND gate for inputs X and Z. **d,e,** An XOR gate built with only two M-Transistors, with an experimental demonstration of the thermal pathway in four states. The symbol ↔ represents "exclusive disjunction."

# Article

For a 1-bit input, the M-Transistor could be configured to function as a NOT gate, as the output thermal signal is the opposite of the input's signal, shortly $T_{out} = \neg\, T_{in}$ (Fig. 2a,b). Similarly, we can build a buffer (a.k.a. YES gate) by exchanging the two constant sources (ii) and (iii) (details in Supplementary Materials "Other logic gates"). The 2-bit logic gates (OR and AND) can also be exhibited in the same M-Transistor without requiring cascaded multiple units. For an OR gate, we select inputs X and Y as the binary inputs $(T_{out} = T_X \cup T_Y)$ while keeping the input Z as $T_H$ (1). Its "disjunction" logic defines that one or both of the alternatives is a high temperature (1); if neither of them is satisfied, we get a low temperature (0) as a result. Constructing an AND gate requires keeping input Y as $T_L$ (0) and selecting inputs X and Z as the inputs $(T_{out} = T_X \cap T_Z)$. Notably, a single M-Transistor switches functions among NOT, YES, OR, and AND logics by simply reconfiguring the sources, leading to much more compact logic circuits than their electronic counterparts. Also, this reprogrammable strategy is distinct to existing designs that cannot be altered once fabricated[6-10,14-16]. In addition, all the above logic functions are persistently reversible without requiring manual resets, which is another key factor for constructing complex networks and has not been successfully achieved in many previous devices[6-8,19-21,32-34].

In conventional computing, complex gates inevitably require the layering of multiple transistors, thus necessitating substantial efforts and dissipation in circuit construction and tuning. Electrical XOR operation cascades ten electrical transistors (five PMOS and five NMOS) or four basic logic gates (one NOT, one OR, and two AND gates). However, we can build NOR, NAND, XOR, and XNOR logic gates by combining only two M-Transistors in series or parallel (Fig. 2d, with more details in Supplementary Materials "Other logic gates"). Because thermal signals are used for both inputs and outputs, M-Transistors can connect and share data with each other, permitting the previous input/output as the subsequent input. An XOR gate (Fig. 2e) is experimentally verified only if two mutually exclusive inputs would result in a high temperature (1), which could enable local summation in arithmetic logic units.

# Article

# Mechanical memory

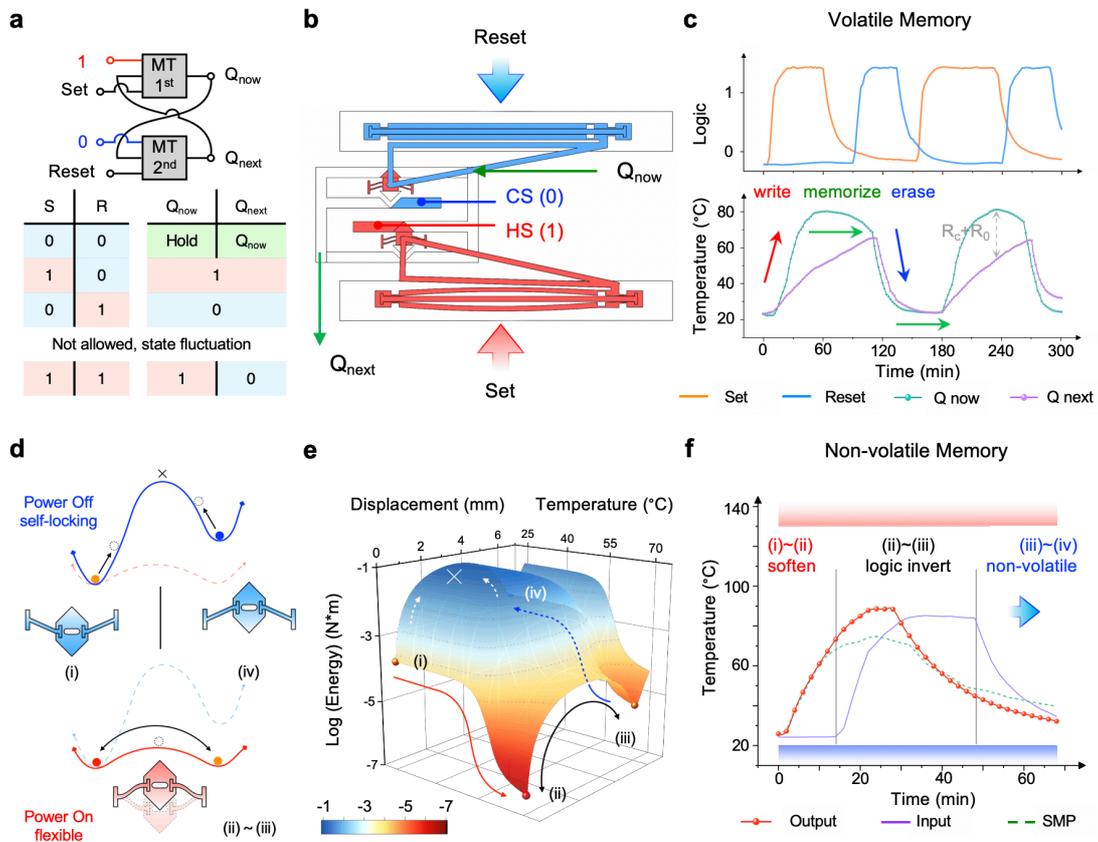

**Fig.3|Rewritable volatile and non-volatile mechanical memories. a,** Principle and transition table of a mechanical register (volatile), consisting of two loop-connected M-Transistors. **b,** Schematic of the mechanical register with "Set" and "Reset" commanding heating and cooling inflows, respectively. **c,** Experimental results of the mechanical register showing a continuous and periodic operation, with erasable and rewritable bit information. **d,** Mechanism of a type of non-volatile memory in which the polymer of the bistable element is replaced with a shape memory polymer (SMP). The self-locking status of the SMP at low temperatures limits the snap-through between (i) and (iv). **e,** Energy portrait of the bistable actuator with SMP, showing that heating softens the SMP; thus, the actuator provides enough energy to overcome the energy barrier between (ii) and (iii). **f,** Experimental results of the non-volatile mechanical memory, which has the same configuration as the inverter (NOT gate) but maintains its previous state after removing energy (compared with **Fig. 1d**).

Data storage is equally important as logic operations in computing architectures, but strict barriers exist to differentiate their functions. While various in-memory computing electronics are proposed to eliminate this barrier[39,40], mechanical memory and logic are currently two separate systems with distinct and incompatible designs. Owing to the reprogrammable properties of M-Transistors, we achieve both mechanical logic and memory units using the same building blocks. Technically, a set-reset latch as a universal register furnishes erasable and rewritable operations of a binary bit[10,11]. The sequential outputs depend not only on its current input but also on its current operation, such as holding previous values or redefining values (Fig. 3a). Interestingly, two mirrored M-Transistors compose a locally closed loop to circulate current and future outputs $Q_{now}$ & $Q_{next}$ (Fig. 3b). Periodic set-reset experiments verified the repeated registration of mechanical bits as shown in Fig. 3c. After the "set" signal (orange line) reaches logic "1", the output heats up and is considered as "write". Only when the "reset" signal (blue line) kicks in to "1", the output rapidly cools down, considered as "erase" (details in Supplementary Video 4 "Erasable, rewritable mechanical register"). We note that transmission delays and losses can be reduced by optimizing the contact ($R_c$) and assembly ($R_0$) thermal resistances. Besides, continuous operation over 1000 min also further verifies the stability and durability of the mechanical latch (see Supplementary Video 5, "Durability tests of mechanical transistors").

Neuromorphic computing typically requires non-volatile memories[41,42], which could be achieved by prominent and sustained mechanical bistability, even after completely losing energy sources. To transform the M-Transistor into a non-volatile device, we harness the self-locking mechanism by replacing the materials of the soft bistable beams with a shape memory polymer (SMP), namely polylactic acid (Fig. 3d). Specifically, the stiffness and energy barrier of SMP beams are closely related to temperatures, and thus the metamaterial sensor could only trigger the bistable actuator when the SMP exceeds its glass transition temperature ($T_g$~60°C); Otherwise, the self-locking of the SMP bistable element prevents snap-through, creating a non-volatile operation. We use energy portrait to further illustrate the energy evolution of the SMP element (Fig. 3e). While softening the SMP with heat, the two stable states of the bistability could be switched in the low-energy channel, resulting in an inverter similar to the dynamic behavior shown in Fig. 1b-d. On the other hand, the energy barrier between the two stable states significantly increases when the SMP actuator is cooled to room temperature, causing the actuator to lock to its position as the sensor's thermomechanical force does not suffice to overcome the energy barrier, thus maintaining long-term memory. This effect is experimentally shown as the output stays at "0" while the input switched from "1" to "0", which is clearly opposite to the inverter behavior in Fig. 1b-d (Fig. 3f, and details in Supplementary Video 6 "Non-volatile mechanical memory"). We note that a power outage does not affect the non-volatile state; only if we reheat the SMP, the bistable actuator can return to its original shape as a bit rewriting operation. These dynamic and non-volatile memories promote a general logic-with-memory mechanical computing architecture.

# Article

# Computing architecture

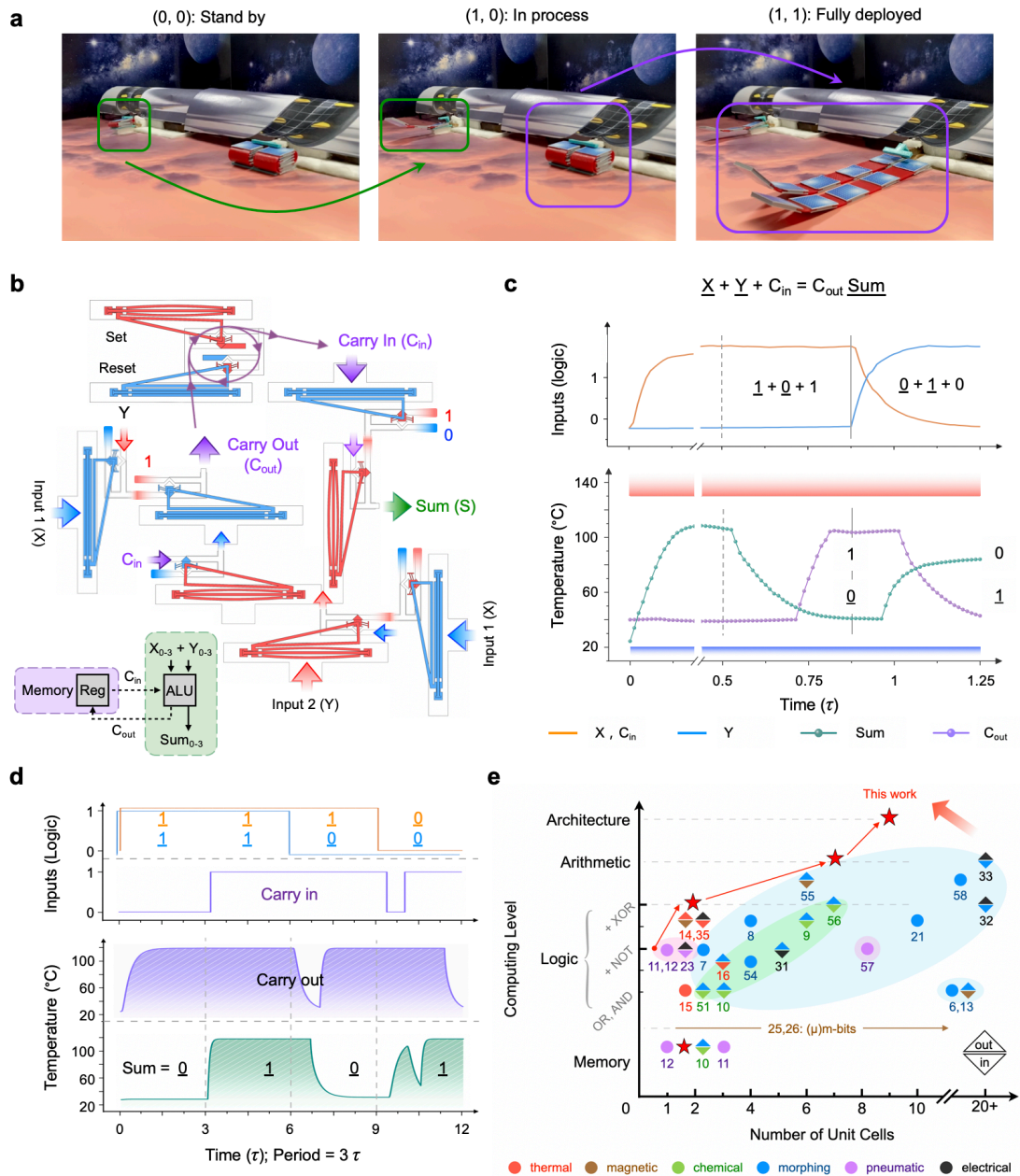

**Fig.4 | Computing architectures with M-Transistors for aerospace applications and logic-with-memory computing. a,** Unfolding deployment of aerospace solar sails due to sequential logic functions utilizing environmental heat inputs. **b,** Implementation of a logic-with-memory mechanical calculator. The ALU (full adder) comprises seven M-Transistors for arithmetic logic and the Reg (set-reset latch) for periodically registering the "carry" bit. **c,** Experimental results of the mechanical full adder. M-Transistors are compatible with sequential and combinatorial logic, firstly caching data, then operating "1+0" with carry "1"; finally inverting all bytes. **d,** Thermal simulation of a four-bit

processing network. We take X: 0011 + Y: 0111 = Sum: 1010 as an example, accompanied by iterative carry. The "carry out" of the previous period is transferred to the "carry in" of the next period. Notably, the disturbance at 6τ is self-corrected in the end. **e,** Benchmarking cutting-edge mechanical computing. We present a comprehensive summary of existing mechanical computing, highlighting our intelligent architecture with M-Transistors and the logic-with-memory architecture.

The main goal of modern mechanical computing is to compensate for electronic malfunctions in harsh conditions, such as aerospace, nuclear, and polar environments. Here, we demonstrate that natural heat sequentially deploys folded solar panels with embedded mechanical logic gates (Fig. 4a). In one scenario, a Mars rover experiences power loss and goes to hibernate due to neither external solar power nor internally reserved energy (0, 0). Later, because of heating by reappeared solar power, the thermally responsive clamps first released a small solar panel as a result of the OR logic (1, 0). However, the main solar panel is still folded because catastrophes like Martian sandstorms can pollute and damage large-area panels, and thus it waits for mechanical logic gates to make further decisions. The main solar panel is only allowed to fully deploy only when the small solar panel harvests sufficient energy under a calm weather and provides a heated second input due to the result of an AND logic (1, 1) (details in Supplementary Video 7 "Thermomechanical computing for aerospace"). This is the first demonstration of self-implementation of mechanical logic driven by environmental heat.

More importantly, we create a modular physical architecture with very few M-Transistors for logic-with-memory computing. Taking an electronic full adder as an example, 38 electronic transistors (20 MOSFETs for two XOR gates, 12 MOSFETs for two AND gates, 6 MOSFETs for one OR gate) barely meet the demand for 1-bit "sum" and "carry" as outputs. In contrast, by leveraging the reprogrammability of our platform, only seven M-Transistors are sufficient to perform the equivalent arithmetic function, with two additional M-Transistors to periodically store the "carry" bit for the next computing period (Fig. 4b). With such a configuration, we have experimentally revealed the time-domain signals of a mechanical full adder (Fig. 4c). We define a dimensionless time constant ($\tau$=160 min) as the time scale based on thermal modeling for the heat transfer between two neighboring M-Transistors approaching a steady state (details in Supplementary Materials "Heat transfer modeling"). The M-Transistors within the

# Article

network execute logic in a sequential manner from the periphery to the center because of heat transfer delays. First, the input signals transmit around the periphery (X="carry in"=1, Y=0), and the M-Transistors for the final states "sum" and "carry out" remain at their initial conditions, similar to "caching data" in the central processing unit (CPU). After $0.5\tau$, all the M-Transistors are actuated and implementing designed logic operations, with outputs "sum"=0 and "carry out"=1. Finally, with the updated inputs, the M-Transistor network reboots itself to implement a new instruction with X="carry in"=0 and Y=1 (details in Supplementary Materials "Full adder verification", and Supplementary Videos 8, 9).

Further, the logic-with-memory processing core (Reg + ALU) could persistently calculate the final "sum" using the outputs from previous periods. The spatiotemporal heat transfer of the transistor network was simulated to verify its feasibility (Fig. 4d). As one of the most complex four-digit additions, X: 0011 + Y: 0111, iterative "carry" bit poses a grand challenge to the mechanical computing, which integrating combinational logic and sequential circuits. The blue and orange curves represent the time-domain signals of X and Y, respectively, with the least significant bit being the first value entering the mechanical processing core (entry sequence for X being 1, 1, 0, 0 and Y being 1, 1, 1, 0). Because the first-bit operation does not produce a "carry" bit, the initial instruction that we give to the register is "reset". As a result, all iterative "carry out" can be stored as the signal "1" in the register, and precisely be held for one period. The fourth bit, "carry out," accumulates all the previous hysteresis and delays, and it takes $3\tau$ to reach a steady state. Notably, due to the extra time required for thermal signal propagation and bistable actuation in M-Transistors, a sudden nonideal disruption at the beginning of the third cycle shows up, but it does not affect subsequent calculations as the mechanical network has self-correcting stability. The purple and green areas convey "carry" and "sum" in the adder architecture, and a correct sum of 1010 is reached in the end. We note that, on the one hand, the significant thermal inertia in M-Transistors is the leading cause of delay, and on the other hand, the insensitivity to short-term disturbance ensures excellent robustness. Finally, owing to a lack of metrics for assessing different designs of mechanical computing, we benchmark all recent studies against the functional completeness and the required number of units (Fig. 4e). This work not only shows complete functions of logic and memory but also achieves an integration of them for a

# Article

logic-with-memory computing architecture, with significantly fewer number of units as well.

**Article**

# Outlook

According to the equivalent electronic circuits, we have verified all eight logic gates and designed essential computing circuits, such as a full adder, an SR register, and a non-volatile memory, with a tremendous decrease in the number of transistors. Second, the inevitable heat dissipation and conduction loss are the main limitations of a further larger scale of integration, which can be improved by miniaturization to the mesoscale or performing all the computing in a vacuum environment, as in real space. On the other hand, this delay also allows an insensitivity to unexpected disturbances, as the computation will not be interrupted even if the heating energy is missing for a short period. Third, although the miniaturization of the M-Transistor is possible[43,44] for a faster speed and a higher integration, real-world applications[45,46] may not easily provide a small-scale alternative hot and cold temperature distribution and could also accommodate macroscopic thermal devices, such as nuclear facilities, space exploration, and intelligent buildings.

In summary, as a new paradigm of mechanical computing, our concept of a fundamental mechanical transistor exhibits its versatility and indispensability. Compared with existing mechanical logic gates, we highlight the emergence of "mechanical logic-with-memory computing" and corresponding combinational (logic) and sequential (memory) circuits with full functional completeness. We anticipate that our design would be a milestone for more sophisticated physical intelligence and may inspire engineering in broader fields such as integrated logic and memory in molecular computing or living organisms[47,48].

# Article

# Article

# Article

## Materials and Methods

### Kinematic models of the CPS

We define input and output mechanical signals as $D_{in}$ and $D_{out}$, respectively. We provided an analytical expression of the thermomechanical response of a compliant porous structure (CPS) with compliant hinges whose main loading mode is pure bending (see Supplementary Materials "Thermomechanical deformation of Compliant Porous Structure (CPS)").

The input displacement $D_{in}$ is calculated as

$$D_{in} = L_1 + L_2 + 2L_h - L_3' \tag{1}$$

The output displacement $D_{out}$ is determined as

$$D_{out} = 2\sqrt{\left(\frac{3L_h}{2} + L_1 + L_2\right)^2 + \left(t + G_0 - \frac{t_h}{2}\right)^2} \sin\frac{\theta_1}{2} \tag{2}$$

, where $L_i$ and $t_h$ represent the length and thickness of hinges, respectively, and $\theta_i$ (i=1-3, h) is the angle of the curve, as shown in Figure S1. The amplification ratio $k = D_{out}/D_{in}$ increased with a decrease in both $L_1$ and $D_{in}$. The maximum application ratio $k_{max}(= 47.4)$ can be obtained with $L_1 = 10mm$ and $D_{in} = 0.1mm$, as shown in Figure S2-S4.

### Spring modeling of the bistable element

To better understand the physics behind the snap-through behavior, we develop an analytical model of a bistable structure based on a soft spring model[49,50]. Finally, we can obtain the force-displacement relationship of the system as

$$F = k_1 L\left[1 - \sqrt{1 - 2\sin\alpha\frac{d}{L} + \left(\frac{d}{L}\right)^2}\right]\sin\left[\tan^{-1}\left(\tan\alpha - \frac{d}{L\cos\alpha}\right)\right] +$$

$$k_2\left[\alpha - \tan^{-1}\left(\tan\alpha - \frac{d}{L\cos\alpha}\right)\right]\frac{1}{L\cos\alpha}\frac{1}{1+\left(\tan\alpha - \frac{d}{L\cos\alpha}\right)^2} \tag{3}$$

where $k_{1,2}$ indicate linear spring coefficients. Figures S5-S7 describe the initial geometry of the spring system, including the spring's length $L$, and the inclined angle $\alpha$. An external vertical force F results in a vertical displacement $d$ (see Supplementary Materials "Bistable modeling").

**Design and fabrication of M-Transistors**

The mechanical metamaterial sensor comprises a Kirigami-inspired CPS and an I-shape invar bar. The two parts are designed in CAD software SolidWorks 2019 (Dassault Systèmes SE, France). We used polycarbonate (PC) as the main material for CPS, which is sliced at 100% infill density by Ultimaker Cura (Ultimaker B.V., USA) to generate a Gcode file for 3D printing. Besides, the supports of the CPS and the bistable structure are also made from Ultimaker polycarbonate through 3D printing. CPS can function as a displacement amplifier by embedding a waterjet-cut invar bar (type 4J36 with 60% iron, 32%-36% nickel, Mingshang Metal, China) under a certain temperature difference. The alloy's density and coefficient of thermal expansion are $8.1\ g/cm^3$ and $1 \times 10^{-6}/K$, respectively. The details of the parameters are shown in Supplementary Materials "Material properties."

The bistable soft actuator comprises a photosensitive resin element and two triangular copper prisms. The soft matter is fabricated using stereolithography (Formlabs Inc., USA) or direct ink writing[6,51]. Cu (type T2 with 99.91% Cu, Guanye Co., Ltd., China). was manufactured and polished using computerized numerical control (CNC). We replaced the soft beam with a 3D-printed polylactic acid (PLA) element for the non-volatile memory unit. To avoid the need for additional energy, the main body of the bistable actuator adopted a hollow copper block, connecting heat or cold sources to the PLA element.

**Assembly and heat transfer optimizations**

To reduce the heat loss and achieve a higher temperature output, we cored M-Transistors with aerogels (type JN650, Hegao Materials, China). The density and coefficient of heat conductivity are $0.18 - 0.20\ g/cm^3$ and $0.018\ W/m \cdot K$, respectively. The 8mm-thick aerogel was the base of M-Transistors; 2mm-thick strips are covered on the surface of

# Article

the copper at the output. A liquid metal[52] with good thermal conductivity was smeared on the interfaces to fill micro air gaps, thus reducing the thermal contact resistance. The liquid metal mainly consists of gallium and indium (Yelengxing, China), which can keep its properties stable at -50−150°C. The density and the coefficient of heat conductivity are approximately 5.8 $g/cm^3$ and $140\,W/m \cdot K$, respectively. Three kinds of glue were used during the assembly, including the cyanoacrylate adhesive, Teflon tape, and Loctite EA E-120HP adhesive. A cyanoacrylate adhesive was applied to connect the bistable element, its support, and the aerogel base. Teflon tape was used to adhere the thermocouple and heated ceramic to the copper. The Loctite EA E-120HP adhesive with good heat resistance can maintain stickiness under high-temperature conditions, including attaching a heated ceramic, water-cooling head, copper, and aerogel.

**Thermal resistance models of M-Transistors**

To verify the logic operation of the M-Transistor, we use a heat transfer model and analyze an inverter (NOT gate) as an example.

$$T_{out\ (in=0)} - T_{out\ (in=1)} = (T_H - T_C) \cdot \frac{R_{air} - R_{contact}}{\sum_{i=1}^{4} R_i + R_c + R_{air}} \gg 0 \tag{4}$$

, where $T_H$ and $T_C$ represent the temperatures of the heat and cold sources, respectively. $R_i\ (i = 1, 2, 3, and\ 4)$ denotes the thermal resistance of copper (see Supplementary Materials "Heat transfer modeling of M-Transistors"); $R_{contact}$ represents the contact thermal resistance, and $R_{air}$ indicates the thermal resistance of air (relatively large, can be regarded as an open circuit).

Based on our previous work[38], the contact thermal resistance is described through an elastic model:

$$R_{contact} = 1.25 \frac{\mu}{\sigma} \cdot \frac{\kappa_{Cu}\kappa_{LM}}{\kappa_{Cu} + \kappa_{LM}} \cdot \left(\frac{\sqrt{2}P}{\mu E}\right)^{0.95} \tag{5}$$

$$E = \left(\frac{1-v_{Cu}^2}{E_{Cu}} + \frac{1-v_{LM}^2}{E_{LM}}\right)^{-1} \tag{6}$$

,where $\mu = \sqrt{\mu_{Cu}^2 + \mu_{LM}^2}$ indicates surface profile, $\sigma = \sqrt{\sigma_{Cu}^2 + \sigma_{LM}^2}$ is surface roughness, $\kappa_{Cu;LM}$ and $\nu_{Cu;LM}$ are thermal conductivities and Poisson's ratios of copper block and liquid metal, respectively. $E$ is the effective Young's modulus, and the mechanical sensor with an asymmetric Kirigami structure generates a contact pressure $P$ with high-temperature input signals. We adopt numerical fitting with experimental data to quantify contact thermal resistance, $R_{contact}$ changes from 0.8 to $1.5 \times 10^{-3}$ (K·m²/W), as the contact becomes increasingly tight with the increasing input temperature.

**Dimensionless time constant**

We define dimensionless time to guide possible miniaturization and material selection for computing systems. The relaxing time $\tau$ (systematic time constant) represents the time it takes to reduce the temperature difference to $e^{-1}$ of the initial state. Considering the time required to heat the copper via a single M-Transistor:

$$\frac{(T-T_H)}{(T_{initial}-T_H)} = \exp\left(-\frac{\tau}{\rho c_p V R}\right) = e^{-1} \tag{7}$$

, where $\rho$ = 8978 (kg/m³) is copper's density, $c_p$ = 381 (J/kg·K) indicates the heat capacity of copper, and $V$ indicates the volume of the copper block. By fitting the experimental results and thermal simulations, we obtain $\tau$ = 160 min.

**Article**

## Data availability

The codes and source data supporting the findings of this study are available from the corresponding authors upon reasonable request.

## Author Contributions

H.C., C.S., and J.J. conceived the research contents. H.C. and J.W. fabricated the devices and conducted experiments. H.C. analyzed the data and produced the figures and videos. C.S. and B.Z. performed mechanical models and tested material properties. H.C., Z.Z., and Z.W. built thermal models. A.Z., and Y.C. contributed to the background and new concepts of electrical computing. H.C. and J.J. wrote the manuscript. W.Z., L.S., and J.J. reviewed the manuscript. L.S. and J.J. supervised the project. All authors contributed to the discussion and editing. H.C. and C.S. equally contributed to this work.

## Acknowledgements

This research is supported by the National Natural Science Foundation of China (Grants no. 12272225 (J.J.), 12002201 (L.S.)), the Ministry of Science and Technology in China (Grant no. SQ2022YFE010363 (J.J.)), and the Research Incentive Program of Recruited Non-Chinese Foreign Faculty by Shanghai Jiao Tong University (J.J.). We also thank Z. Hu and Z. Yang (Shanghai Jiao Tong University) for the miniaturized design; J. Fan and H. Bao (Shanghai Jiao Tong University) for equipment support; Z. Ren (Max Planck Institute), D. J. Preston (Rice University), W. Li (Tongji University), W. Wang, and Q. Liu (Shanghai Jiao Tong University) for technical discussions.

## Competing interests

The authors declare no competing interests.